\documentclass{aa}

\usepackage{graphicx}
\usepackage{caption}
\usepackage{subcaption}

\usepackage{amssymb,amsmath}
\usepackage{color}
\usepackage{epstopdf}
\usepackage{natbib}
\usepackage{multirow}
\usepackage[colorlinks=true, linkcolor=blue, citecolor=blue, filecolor=blue, urlcolor=blue]{hyperref}
\usepackage{placeins}
\usepackage{txfonts}
\usepackage{soul}
\usepackage{ulem}
\usepackage{orcidlink}

\title{The connection between surface brightness and satellite systems for central galaxies through Illustris TNG}

\author{Silvio Rodriguez\inst{1}\thanks{E-mail:silvio.rodriguez@unc.edu.ar}\orcidlink{0000-0001-5329-5242},
Yamila Yaryura\inst{1,2}\orcidlink{0000-0002-5545-1322},
Jose A. Benavides\inst{3}\orcidlink{0000-0003-1896-0424},
Diego Garcia Lambas\inst{1,2}\orcidlink{0000-0002-2929-4685},
Susana Pedrosa\inst{4}\orcidlink{0000-0001-8249-4435},
Laura D. Baravalle\inst{2}\orcidlink{0000-0001-6163-8807},
Laura Ceccarelli\inst{1,2}\orcidlink{0000-0002-2136-2591},
Heliana E. Luparello\inst{2}\orcidlink{0000-0002-1114-9578},
Lucas Bignone\inst{4}\orcidlink{0000-0002-4746-1627},
Gaspar Galaz\inst{5}\orcidlink{0000-0002-8835-0739}}
\authorrunning{S. Rodr\'iguez et al.}
\institute{
Observatorio Astron\'omico de C\'ordoba, Universidad Nacional de C\'ordoba, X5000BGR, Argentina
\and
Instituto de Astronom\'ia Te\'orica y Experimental, UNC-CONICET, C\'ordoba, X5000BGR, Argentina
\and
Department of Physics and Astronomy, University of California, Riverside, CA, 92521, USA
\and
Instituto de Astronomía y F\'isica del Espacio, CONICET-UBA, 1428 Buenos Aires, Argentina.
\and
Instituto de Astrof\'isica, Pontificia Universidad Cat\'olica de Chile, Vicu\~na Mackenna 4860, 7820436 Macul, Santiago, Chile.
}

\date{Accepted --. Received --; in original form --}

\abstract{We analyse different properties of central low-surface-brightness galaxies (LSBGs) and their satellite systems using the simulation Illustris TNG-100, in order to deepen our understanding of the formation mechanism of LSBGs in a $\Lambda$CDM cosmology. We find differences in the spin and the concentrations of the LSBGs haloes and the host haloes of high-surface-brightness galaxies (HSBGs), consistent with previous studies. By analysing their spatial and kinematical distribution of satellites, we find that LSBGs tend to have a larger number of satellites than HSBGs and with a larger velocity dispersion. Moreover, we obtain a continuous relation between the number of satellites and surface brightness, particularly for massive central galaxies. We also find a relation between surface brightness and the relative tangential velocity of the satellites. For a given stellar mass, the existence of LSBGs is strongly correlated with their satellite system dominated by rotation. Furthermore, the satellite system is systematically in counter-rotation with respect to the primary disc in LSBGs. We propose that this fact reflects that these galaxies have not experienced a significantly high rate of mergers, which are more likely associated with radial orbits expected in systems of galaxies with a high surface brightness.
}


\begin{document}

\maketitle

\section{Introduction}
\label{sec_intro}

Observationally, low-surface-brightness (LSBGs) studied at the Local Universe display distinctive structural and baryonic properties, which often contrast with those of high-surface-brightness galaxies (HSBGs). Their stellar discs tend to be unusually extended, with large effective radii and shallow radial profiles, while their gas reservoirs can be massive, metal-poor, and inefficient in forming stars \citep{Kuzio2004,wyder2009,Rong2020,Schombert2021}. These properties contribute to long depletion timescales and a more gradual evolution  when compared with more compact systems. Although LSBGs are commonly associated with low-density regions \citep{Mo1994, Galaz2011, Perez-M2019}, recent studies emphasize that their distribution is not uniform, with environmental signatures emerging in satellite populations, tidal features, and local dynamics \citep{Rosenbaum2009,muller2019}. Dynamical analyses reveal that LSBGs are often dark matter dominated even within their optical radii \citep{deBlok1997}, though several massive examples show baryon-dominated inner parts, leading to a broader diversity than previously expected \citep{lelli2010,saburova2021}. Furthermore, the specific angular momentum of LSBGs tends to be systematically lower than that of HSBGs at fixed stellar mass \citep{Salinas2021, Perez-M2022, Zhu2023, Pallero2025}, reinforcing that their formation histories differ significantly.

The theoretical understanding of LSBG formation has advanced considerably over the last decade, driven by analytical models and high-resolution hydrodynamical simulations. A widely discussed scenario proposes that LSBGs reside in dark matter haloes with unusually high spin parameters, which naturally lead to diffuse stellar distributions when baryons inherit this angular momentum  \citep{Kim2013,kulier2020}. Cosmological simulations support this connection, showing that high spin haloes yield galaxies with systematically lower surface brightness at fixed stellar mass \citep{Perez-M2022, Perez-M2024}. However, halo spin alone cannot account for the entire diversity of LSBGs. Additional mechanisms identified in cosmological and zoom-in simulations include strong stellar or AGN feedback episodes that expand the stellar and gas components \citep{Chan2018,DiCintio2019}, gas-rich or coplanar mergers that preserve rotational support and build extended discs \citep{wright2021}, prolonged accretion of cold, high-angular-momentum gas \citep{saburova2021}, or multiple evolutionary scenarios where internal and external processes jointly shape diffuse systems \citep{Papastergis2017,Wright2025,Wu2025}. Various simulation suites – including NIHAO \citep{Wang2015}, Horizon-AGN \citep{Dubois2014}, EAGLE \citep{Schaye2015}, and Illustris TNG \citep{tng2} have successfully reproduced the observed sizes, colours, and kinematics of LSBGs, indicating that several physical channels can give rise to low-surface-brightness structures \citep[e.g.][]{Martin2019,kulier2020,Perez-M2022,Benavides2023, Benavides2024,Rong2024}.

Although significant progress has been made, a key aspect of galaxy evolution that remains relatively unexplored in the context of LSBGs is the role played by their satellite systems. The kinematics, abundance, and spatial distribution of satellites encode detailed information about halo assembly, merger histories, and angular momentum acquisition. Numerous studies have demonstrated that satellites can trace the spin and dynamical state of their host haloes \citep{Sales2010,Sales2012,Rodriguez2021}, and that anisotropic or radial accretion tends to heat or perturb discs, potentially increasing their central surface brightness and promoting more compact configurations \citep{Tang2024,Nagesh2024}. Observations and simulations further show that co-rotation or counter-rotation between satellites and their host discs carries essential information about the directionality and frequency of past accretion events \citep{Pallero2025, Wright2025}. These findings suggest that satellite dynamics should be intimately related to the emergence of both LSBGs and HSBGs, particularly given the strong dependence of disc structure on angular momentum transfer. In addition, the abundance of satellites itself is shaped by environmental conditions and halo assembly, as suggested by observations and simulations of both diffuse and high-surface-brightness systems \citep{Rosenbaum2009,muller2019,Benavides2023}.

Despite the relevance of these processes, a systematic comparison of the satellite systems of LSBGs and HSBGs at fixed stellar mass has not been carried out. In particular, the connection between satellite orbital structure, halo spin, and central surface brightness remains poorly constrained. Understanding this connection is crucial, as satellite accretion provides a principal mechanism for both mass growth and angular momentum redistribution in galaxies, and may play a decisive role in determining whether a system evolves into a diffuse, extended LSBG or into a more concentrated HSBG.

As mentioned before, galaxy haloes are key ingredients of galaxy formation and evolution models. Among the many topics of interest on the relation between haloes and galaxies is the way galaxies populate the haloes \citep[e.g.][]{Rodriguez2021}.
Besides the satellite to central abundance relation, there could be other interesting and related issues which connect the dynamics of satellites and their possible dependence on the surface brightness of central galaxies,
as well as geometrical and astrophysical relations. In fact, the continuous accretion of sub-haloes, both luminous (satellites) and dark, generates streams of infalling stars and dark matter that can be incorporated onto the external disc in the outskirts of galaxies, inducing changes in the galaxy's effective surface brightness. This would contrast with a more violent scenario where direct mergers could enhance the light concentration of the galaxy. Thus, the remaining satellite system and its astrophysical and dynamical properties could in principle be significantly correlated with the primary surface brightness.
In this work we focus on the study of spatial structure and kinematics of satellite systems as a function of the surface brightness of central galaxies in the Local Universe. Our main goal is to shed light on possible formation scenarios of high and low-surface-brightness galaxies.

This paper is organised as follows. In Sect.~\ref{sec_sample} we describe the samples. In Sect.~\ref{sec_sat} we analyse the  properties of the satellite systems. And finally we present our concluding remarks in Sect.~\ref{sec_disc}.

\section{Sample selection}
\label{sec_sample}

We use data from the Illustris TNG simulation, in particular the TNG100-1 hydrodynamic box \citep{tng1, tng2, tng3, tng4, tng5, Nelson2019}\footnote{\href{}{https://www.tng-project.org/}}. This simulation has a comoving box size of 75000 $\mbox{kpc}\,h^{-1}$, with $1820^3$ dark matter particles with a mass given by 7.5$\times10^6 ~ \mbox{M}_{\sun}$. For the baryon component, the average mass of the gas particles is 1.4$\times10^6 ~ \mbox{M}_{\sun}$. The cosmological parameters of the simulation are consistent with Planck 2015 cosmology \citep{Planck2015}, given by $\Omega_m=0.3089$, $\Omega_{\Lambda}=0.6911$ and $h=0.6774$. 

The halo identification was made with a {\sc FOF} algorithm \citep{Huchra1982, Press1982, Davis1985}, while the substructures (subhaloes) within the halo were detected by the {\sc SUBFIND} algorithm \citep{Springel2001, Dolag2009}. The central galaxy of a halo is defined as the subhalo with the largest number of bound particles and gas cells. The rest of the subhaloes are considered as satellite galaxies. In this work we used the snapshot corresponding to $z=0$.

\subsection{Low and high-surface-brightness samples}
\label{sec_centdef}

\begin{figure}
    \centering
    \includegraphics[width=0.475\textwidth]{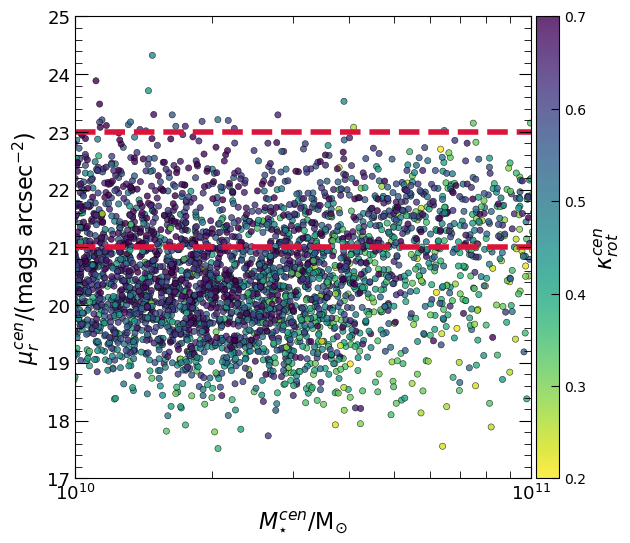}
    \caption{Mean surface brightness ($\mu^{cen}_r$) as a function of the stellar mass ($M^{cen}_{\star}$) for all central galaxies with $M^{cen}_{\star} = 10^{10} - 10^{11} ~ \mbox{M}_{\odot}$, coloured by $\kappa^{cen}_{rot}$. The horizontal dashed red lines indicate the threshold to select high and low-surface-brightness galaxies.}
    \label{fig:all_mus}
\end{figure}

From this simulation we use central galaxies because some properties such as the halo size $R_{200}$ (defined as the radius centred in a halo that comprises a volume with a density that is 200 times the critical density), the halo mass $M_{200}$ (the mass enclosed within $R_{200}$) and the spin parameter \citep[$\lambda$, defined by][]{Bullock2001} are well defined only for central\footnote{Notice that the supra index $cen$ indicates that the quantity corresponds to central galaxies, while the supra index $sat$ is for a quantity measured for satellites.} galaxies. From these galaxies, we select those with a stellar mass ($M^{cen}_{\star}$) in the $10^{10}-10^{11} \mbox{M}_{\sun}$ range, which includes 3110 galaxies. We name this set as the base sample. Along this work we study different samples of central galaxies, all of them based on this data set.

\begin{figure*}
    \centering
    \includegraphics[width=0.32\textwidth]{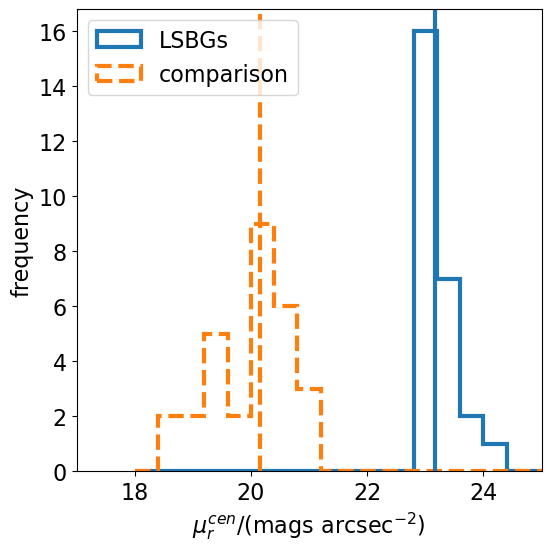}    
    \includegraphics[width=0.32\textwidth]{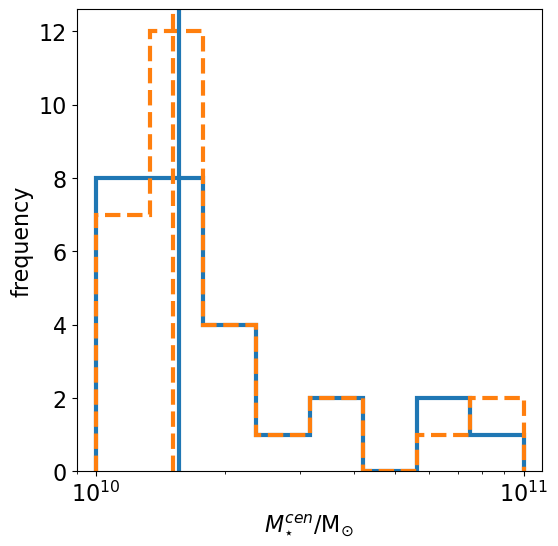}
    \includegraphics[width=0.32\textwidth]{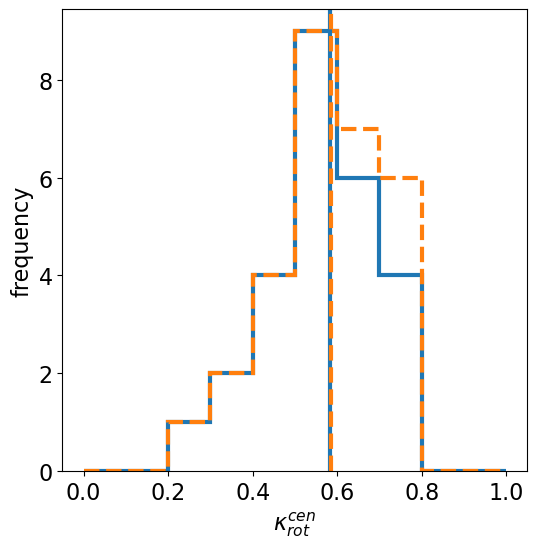}
    \caption{Distributions of the mean surface brightness $\mu^{cen}_r$ (left panel), stellar mass $M^{cen}_{\star}$ (middle panel), and morphology parameter $\kappa^{cen}_{rot}$ (right panel), for LSBGs (solid blue lines) and comparison (dashed orange lines) samples. 
    In each panel, vertical lines indicate the median of each distribution (in their respective colours). }
    \label{fig:basic_dist}
\end{figure*}

\begin{figure*}
    \centering
    \includegraphics[width=0.24\textwidth]{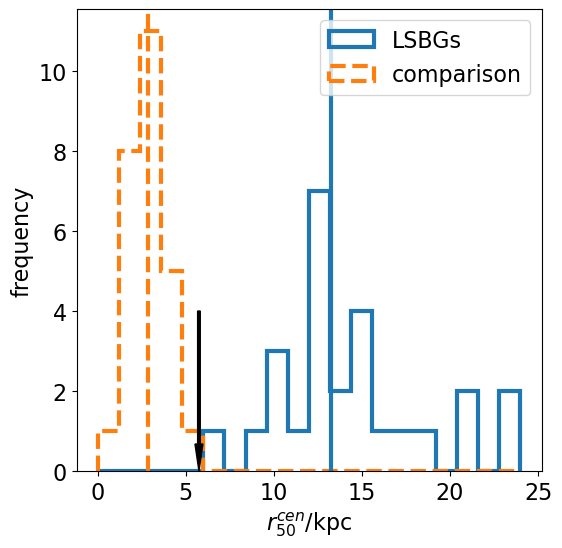}
    \includegraphics[width=0.24\textwidth]{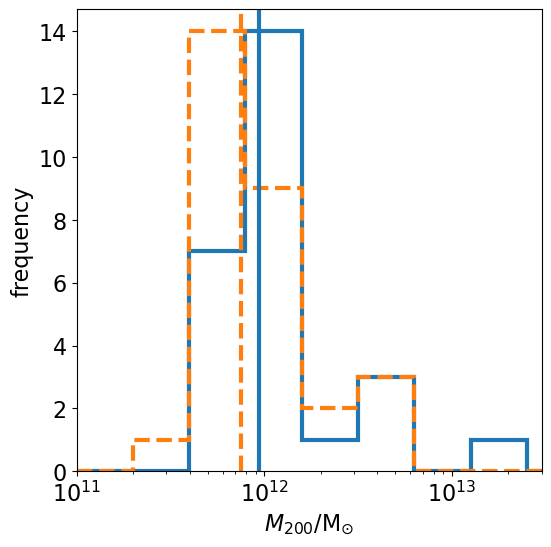}
    \includegraphics[width=0.24\linewidth]{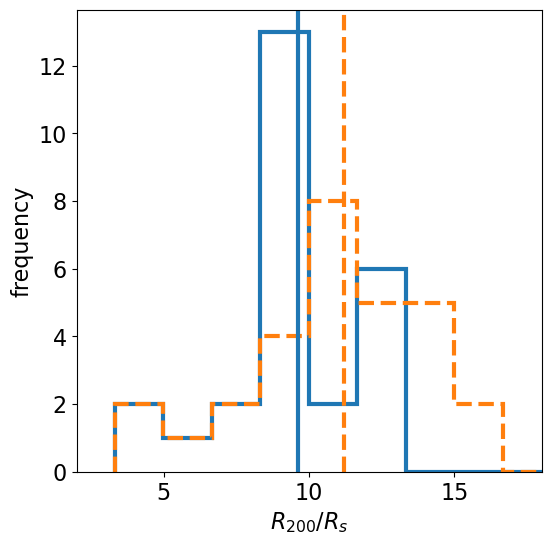}
    \includegraphics[width=0.24\linewidth]{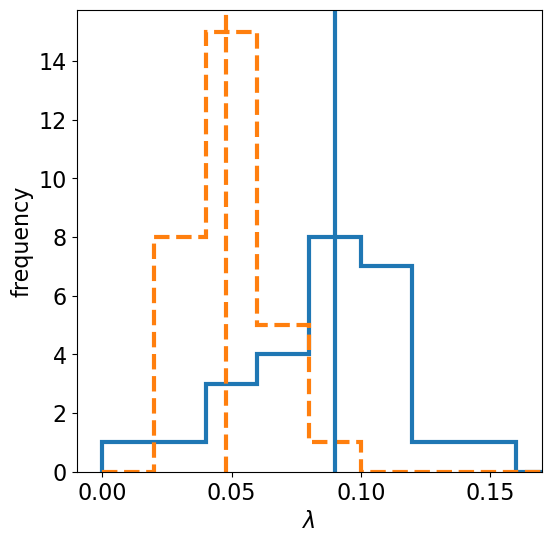}
    \caption{From left to right, distribution for the projected stellar half-mass radius ($r^{cen}_{50}$), halo mass ($M_{200}$), halo concentration ($R_{200}/R_s$), and the dimensionless spin parameter ($\lambda$). The colour code is the same as Fig.~\ref{fig:basic_dist}. In each panel, vertical lines indicate the median of each distribution (in their respective colours), and the black arrow in the first panel indicates the half-light radius of the Milky Way.}
    \label{fig:derived_dist}
\end{figure*}

For the base sample we measure the surface brightness ($\mu^{cen}_{r}$) following the procedure described in \citet{Perez-M2022}, in which $\mu^{cen}_{r}$ is defined by

\begin{equation}
    \mu^{cen}_r = m^{cen}_r + 2.5\log(\pi (r^{cen}_{50})^2)
    \label{eq_mu}
\end{equation}

where $m^{cen}_r$ is the apparent magnitude within the projected half-light radius ($r^{cen}_{50}$). The subindex $r$ indicates that these quantities are measured using the SDSS $r$ band \citep{York2000}.
We also measure $\kappa^{cen}_{rot}$, defined as the ratio between the rotational kinetic and the total kinetic energy of the stellar component of each galaxy \citep{Sales2010}. The $\kappa^{cen}_{rot}$ parameter is a proxy of galaxy morphology as discussed by  \citet{Sales2012} who showed $\kappa^{cen}_{rot}$ to be closely related with stellar circular motions, concluding that the higher the value of $\kappa^{cen}_{rot}$, the most  highly rotationally supported is the galaxy, implying a larger disc component. In Fig.~\ref{fig:all_mus} we show the relation between the $\mu^{cen}_{r}$ parameter as a function of $M^{cen}_{\star}$, coloured by the $\kappa^{cen}_{rot}$ parameter values for each selected galaxy.

From the base sample we select the galaxies with a $\mu^{cen}_r$ value higher than 23 $\mbox{mags\,arcsec}^{-2}$ as our LSBGs sample, with a total of 26 galaxies. We choose this high threshold in order to analyse an extremely low brightness sample. We also select a control sample of HSBGs (hereafter comparison sample), with galaxies with $\mu^{cen}_r$ values lower than 21 $\mbox{mags\,arcsec}^{-2}$, and with similar stellar mass and $\kappa^{cen}_{rot}$ distributions to the LSBGs sample, and a comparable number of galaxies (29). The red horizontal dashed lines in Fig.~\ref{fig:all_mus} show the two $\mu^{cen}_r$ thresholds used to determine the samples. In Fig.~\ref{fig:basic_dist} we show the distributions of $\mu^{cen}_r$ (left), $\kappa^{cen}_{rot}$ (middle), and $M^{cen}_{\star}$ (right) for both samples. Regarding $M^{cen}_{\star}$ and $\kappa^{cen}_{rot}$, their distributions are quite similar, with almost the same median values for both samples. This is confirmed by a Kolmogorov-Smirnov test, which give us a p-value greater than 0.8 for both $M^{cen}_{\star}$ and $\kappa^{cen}_{rot}$, which heavily implies that the underlying distribution is the same in both cases.

\begin{figure*}
    \centering
    \includegraphics[width=0.95\linewidth]{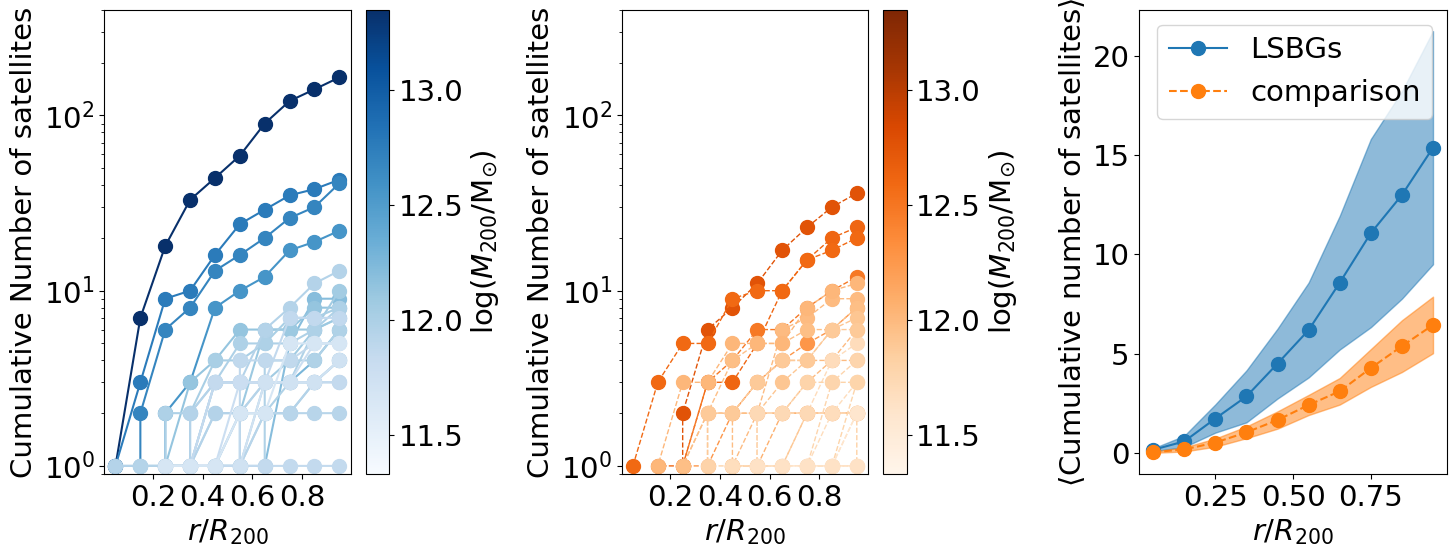}
    \caption{Cumulative number of satellites as a function of the distance to the central galaxy in units of the virial radius ($R_{200}$) of their host halo. The left panel corresponds to the LSBGs sample (blue shaded) the middle panel corresponds to the comparison sample (orange shaded). In both cases, the lines connecting dots correspond to each satellite system for a central galaxy. The colour intensity of each line indicates the mass of the host halo. The right panel shows the mean number of satellites of each sample (in blue, LSBGs and in orange, the comparison sample). The bootstrap errors of the median are indicated by the shadowed areas.}
    \label{fig:numberSat}
\end{figure*}

\begin{figure}
    \centering
    \includegraphics[width=0.95\linewidth]{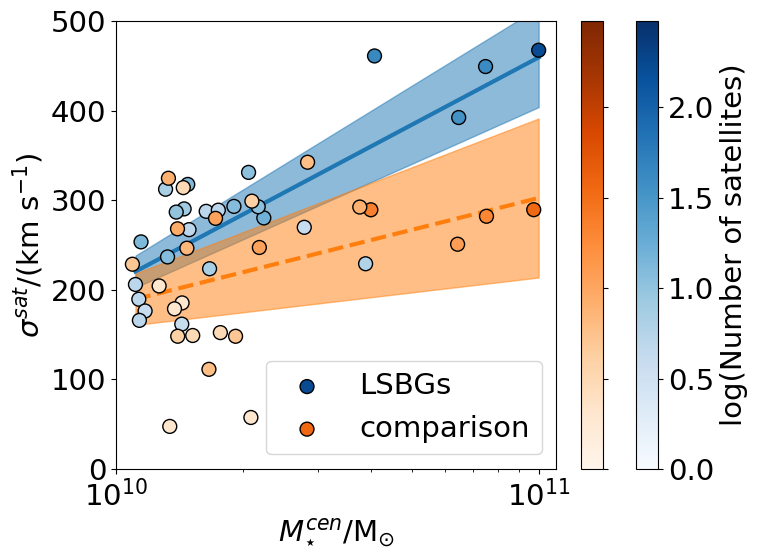}
    \caption{Velocity dispersion ($\sigma^{sat}$) of the satellite system of each host halo as a function of the stellar mass of the central galaxy. The number of satellites is indicated by the shade of the colour for both samples, LSBGs (blue circles) and the comparison sample (orange circles). The solid lines correspond to a linear fit, and the shadowed region indicates the fit error.}
    \label{fig:velocitydisp}
\end{figure}

For a better understanding of the samples, we measure different physical properties of the galaxies and their host haloes. In Fig.~\ref{fig:derived_dist} we show the projected stellar half-mass radius, $M_{200}$ and halo concentration ($R_{200}/R_s$) distributions. Here $R_s$ is the scale radius obtained by fitting a NFW profile \citep{Navarro1997} to the dark matter density profile of the halo \citep[we obtain this value from the catalogue by][]{Anbajagane2022}, and the distribution of the dimensionless spin parameter \citep[$\lambda$, which we calculate following][]{Bullock2001}. As expected, there is an evident difference in the projected sizes of the galaxies in these samples, with LSBGs showing consistently larger sizes. As a reference, the half-light radius of the Milky Way (MW) is $5.75 \pm 0.38 \mbox{kpc}$ \citep{Lian2024},  which is significantly lower than the median value of our LSBGs ($13.23 \pm 0.9 \mbox{kpc}$). We indicate the Milky Way half-light radius by a black arrow in Fig.~\ref{fig:derived_dist}. Despite the fact that the MW and our values are not directly comparable (MW is half-light and ours is half stellar mass), both are suitable estimators of galaxy size. On the other hand, there is no significant difference between the masses of the host haloes among the samples. Notice that the $M_{200}$ distributions are marginally, but noticeably, different, with a tendency of LSBGs to reside in slightly more massive haloes. Notice that $R_{200}/R_s$ values present a slight difference where the haloes that host LSBGs tend to be less concentrated than haloes hosting galaxies from the comparison sample.
According to \citet{Perez-M2024}, there is no difference in concentration related to the surface brightness of the central galaxy, albeit the samples in that paper are defined in a different way than in this work. The $\lambda$ parameter shows a more significant difference between our samples, in which LSBGs tend to live in haloes with high spin. This result is consistent with previous studies \citep[see for instance][]{kulier2020, Perez-M2022, Perez-M2024}.

\subsection{Satellite galaxies}
\label{sec_satdef}

Since we focus on the characteristics of the satellite system around the central galaxies in both of our samples (LSBGs and comparison), it is important to define which satellites we consider as reliable tracers of the structure orbiting those central galaxies.
Firstly, we remove spurious concentrations of mass that are wrongly identified as sub-haloes by {\sc SUBFIND} and are actually part of the central galaxy (as provided in TNG data).
We also apply a mass cut selecting satellites with total mass (dark matter, gas, and stars) equivalent to 100 dark matter particles. This threshold corresponds to a total mass equal to $7.5\times10^8 \mbox{M}_{\sun}$ in TNG-100. Finally, we only consider satellites inside the $R_{200}$ radius of their host haloes. Taking into account these restrictions, we have 400 satellites orbiting around galaxies in the LSBGs sample, and 182 associated to galaxies in the comparison sample.

\section{Analysis and results}
\label{sec_sat}

\subsection{Analysis of LSBG and comparison samples}
\label{sec_satdif}

\begin{figure*}
    \centering
    \includegraphics[width=0.95\textwidth]{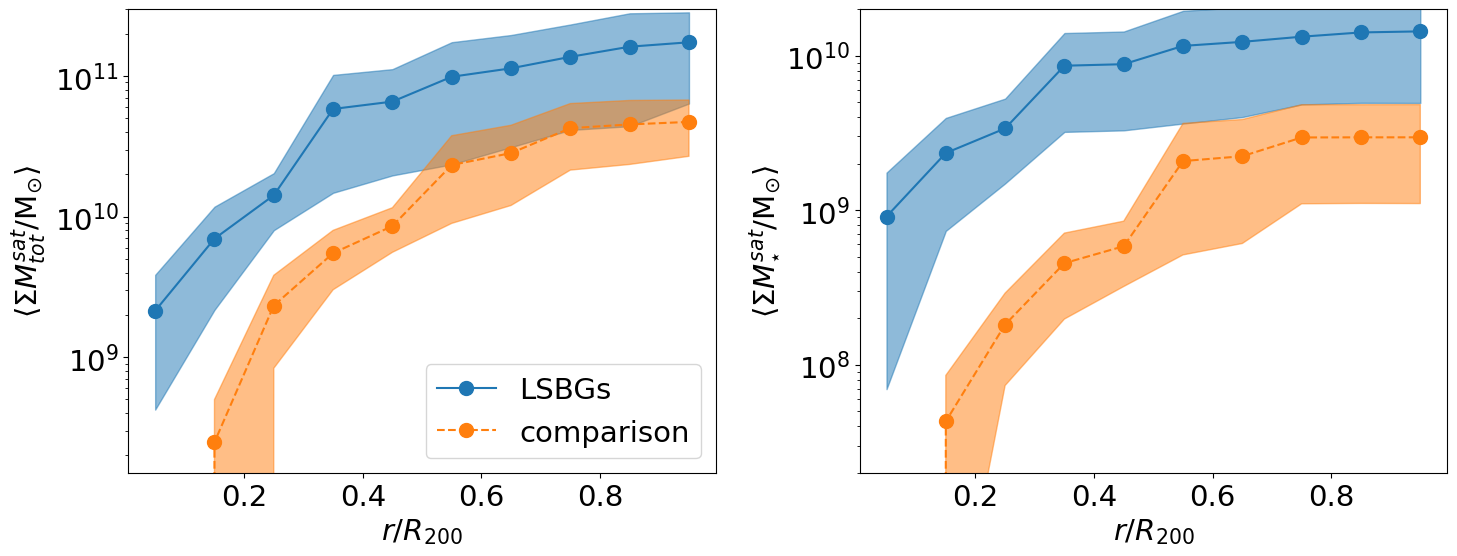}
    \caption{Mean of the cumulative total mass (left panel) and stellar mass (right panel) of satellites as a function of the distance to their central galaxy, in units of $R_{200}$, for the sample of LSBGs (blue) and comparison sample (orange). The shadowed areas indicate the bootstrap error of the mean.} 
    \label{fig:massSat}
\end{figure*}

For our LSBGs and comparison sample (defined in Sect.~\ref{sec_centdef}), we want to compare the number, distribution, and kinematical properties of the satellites (as defined in Sect.~\ref{sec_satdef}) orbiting around these central galaxies. In Fig.~\ref{fig:numberSat} we show the cumulative number of satellites as a function of distance to the central galaxy (in units of $R_{200}$). The left panel shows the satellite system for each central galaxy in the LSBGs sample, with the colour intensity (in blue) indicating the halo mass. The middle panel shows the same statistics for the satellite system around central galaxies in the comparison sample (in oranges). In the right panel we compare the mean number of satellites of both samples. As it can be seen, there is a remarkable difference between the mean abundance of satellites of LSBGs and satellites of galaxies in the comparison sample, particularly for high-mass haloes, which accounts for a factor $\sim 3$ at $R_{200}$. Thus, this result implies either an original deficiency of satellites associated with HSBGs, or otherwise it reflects an evolutionary process associated with mergers with the central galaxy.

The study presented by \citet{Galaz2011} shows that the distance to the 1st and 5th nearest neighbours tend to be closer for HSBGs than for LSBGs. Also \citet{Rosenbaum2009, Galaz2011} find that the number of close neighbours is higher in HSBGs than in LSBGs. Using simulations \citet{Zhu2023} find that the distance to the 5th neighbour is larger in LSBGs compared to all galaxies within $10.2 <log (M_{\star}/M_{\sun}) < 11.6$, but closer compared to a sample of galaxies with similar stellar, dark matter and gas mass to their LSBGs sample. Those results may appear to contradict our results in Fig.~\ref{fig:numberSat}. However, our analysis focuses on low-mass satellites, which is not comparable with the analysis carried out in these works. Also, the sample selection for the neighbours in those works are different from our satellite samples, leading to significantly different spatial scales in their studies, with the conclusion that LSBGs tend to form in less dense environments.

In Fig.~\ref{fig:velocitydisp} we show the satellite system velocity dispersion as a function of the stellar mass of central galaxies for both, LSBG (blues) and comparison (orange) samples. Each dot correspond to the mean velocity dispersion of each satellite system around a central galaxy for both samples. The intensity of the point colours correspond to the number of satellites in each system, as indicated by the colour bar. In this analysis, we only consider satellite systems with at least two satellites. Along with a larger number, satellites of LSBG show a high velocity dispersion at a given central galaxy stellar mass. Also, as can be seen in this figure, there is an apparent strong correlation for LSBGs. In order to quantify the difference in this relation between both samples, we apply a linear fit given by:

\begin{equation}
    \sigma^{sat} = \alpha \times \log(M^{cen}_{\star}/(10^{10} \mbox{M}_{\sun})) + \beta,
    \label{eq:fit}
\end{equation}

where the $\alpha$ and $\beta$ values derived from the fit are in Table~\ref{tab:fit}. This different behaviour of $\sigma^{sat}$ for the LSBG and comparison samples adds an important difference to the previous result on the number of satellites. We argue that at a given stellar mass of a central galaxy, the detected low velocity dispersion of satellites associated to galaxies in the comparison sample can be explained through a stronger dynamical friction, implying a more efficient merging of satellites onto the central. We stress the fact that both LSBGs and comparison samples have the same distribution of host halo mass, implying that the evolution of the baryonic component of the system of satellites and the stellar concentration of the central galaxies are strongly associated. As mentioned in the introduction, the fate of remaining satellites may give key clues on the relation between the primary accretion history and its disc formation.

\begin{table}[]
    \centering
    \caption{Fit parameters for the $\sigma^{sat}-M^{cen}_{\star}$ relation obtained from the linear fit shown in Fig~\ref{fig:velocitydisp}.}
    \begin{tabular}{lcc}
        \hline
         & $\alpha$ & $\beta$ \\
         & $(\mbox{km\,s}^{-1})$ & $(\mbox{km\,s}^{-1})$   \\\hline
         LSBGs       & $250.38\pm39.54$ & $209.08\pm15.96$ \\
         comparison  & $118.35\pm62.86$ & $184.18\pm25.9$  \\\hline
    \end{tabular}
    \label{tab:fit}
\end{table}

Given the results on the number distribution and velocity dispersion of satellites of our LSBGs and comparison samples, we have explored the mass distribution in both stars and total mass (including stars, gas, and dark matter) for the two samples.
For this aim, we have calculated the cumulative mass distributions as a function of distance from the central galaxies in units of $R_{200}$. The results are shown in Fig.~\ref{fig:massSat}, where it can be seen that at distances close to $R_{200}$, both the total mass and the stellar mass converge to a comparable value. However, in the nearest region to the central up to $\sim 1/3 R_{200}$, the LSBGs have significantly larger cumulative fractions of both stellar and total mass. Thus, our findings here reinforce the argument of dynamical friction acting more severely in HSBGs, which may imply a lack of massive satellites at close distances from the central galaxy.

\subsection{Satellite system properties as a function of the central galaxy $\mu^{cen}_{r}$ value }
\label{sec_satcont}

\begin{figure}
    \centering
    \includegraphics[width=0.475\textwidth]{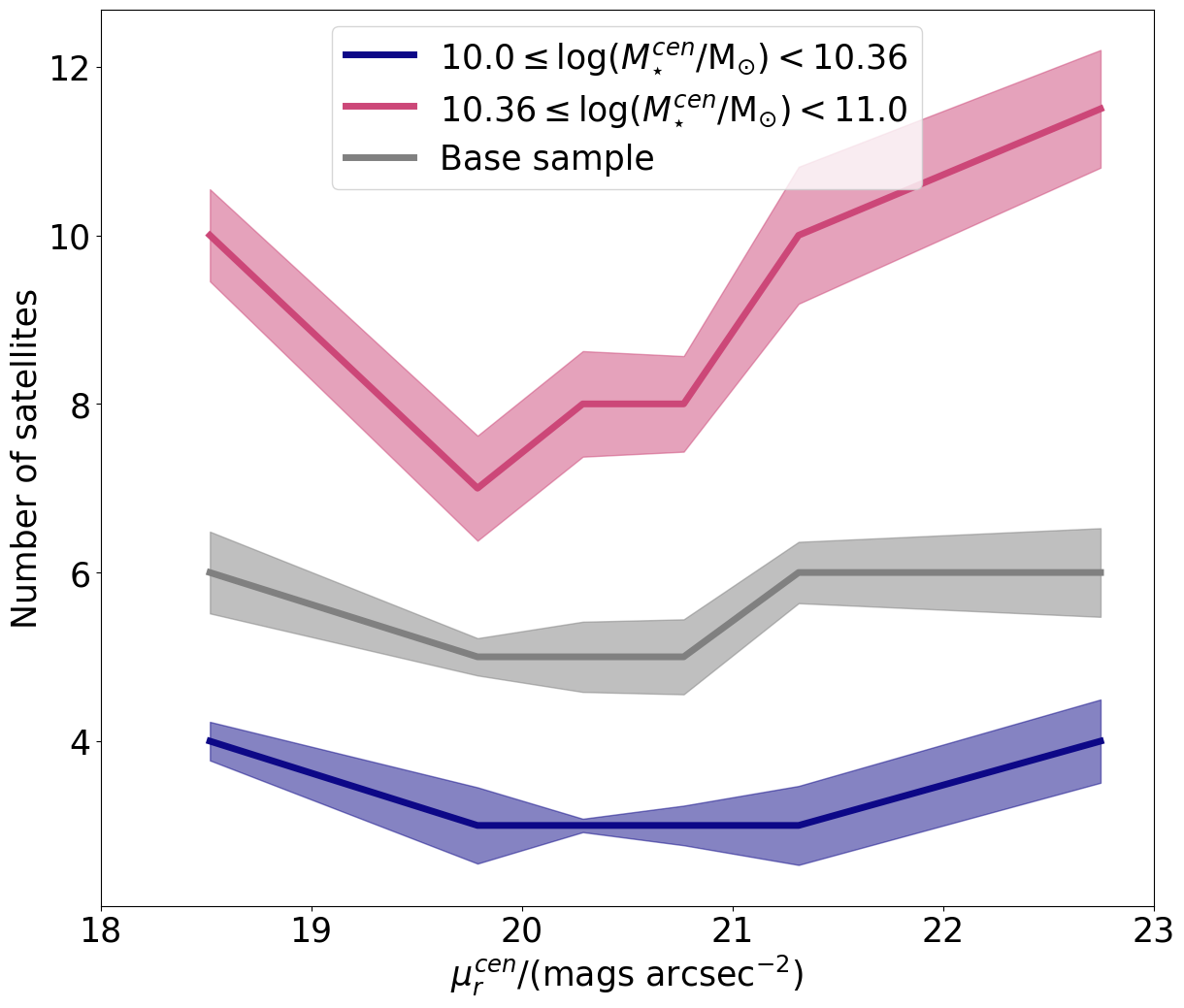}
    \caption{Median number of satellites as a function of the surface brightness ($\mu^{cen}_r$) for all central galaxies in the mass range analysed, split by stellar mass in low-mass ($10 \leq \log(M^{cen}_{\star} / \rm{M_{\odot}}) < 10.36$, indigo line) and high-mass ($10.36 \leq \log(M^{cen}_{\star} / \rm{M_{\odot}}) \leq 11.0$, pink line), while the base sample is presented in grey. The shadowed areas indicate the median bootstrap error.} 
    \label{fig:number_cont}
\end{figure}

\begin{figure*}
    \centering
    \includegraphics[width=0.475\textwidth]{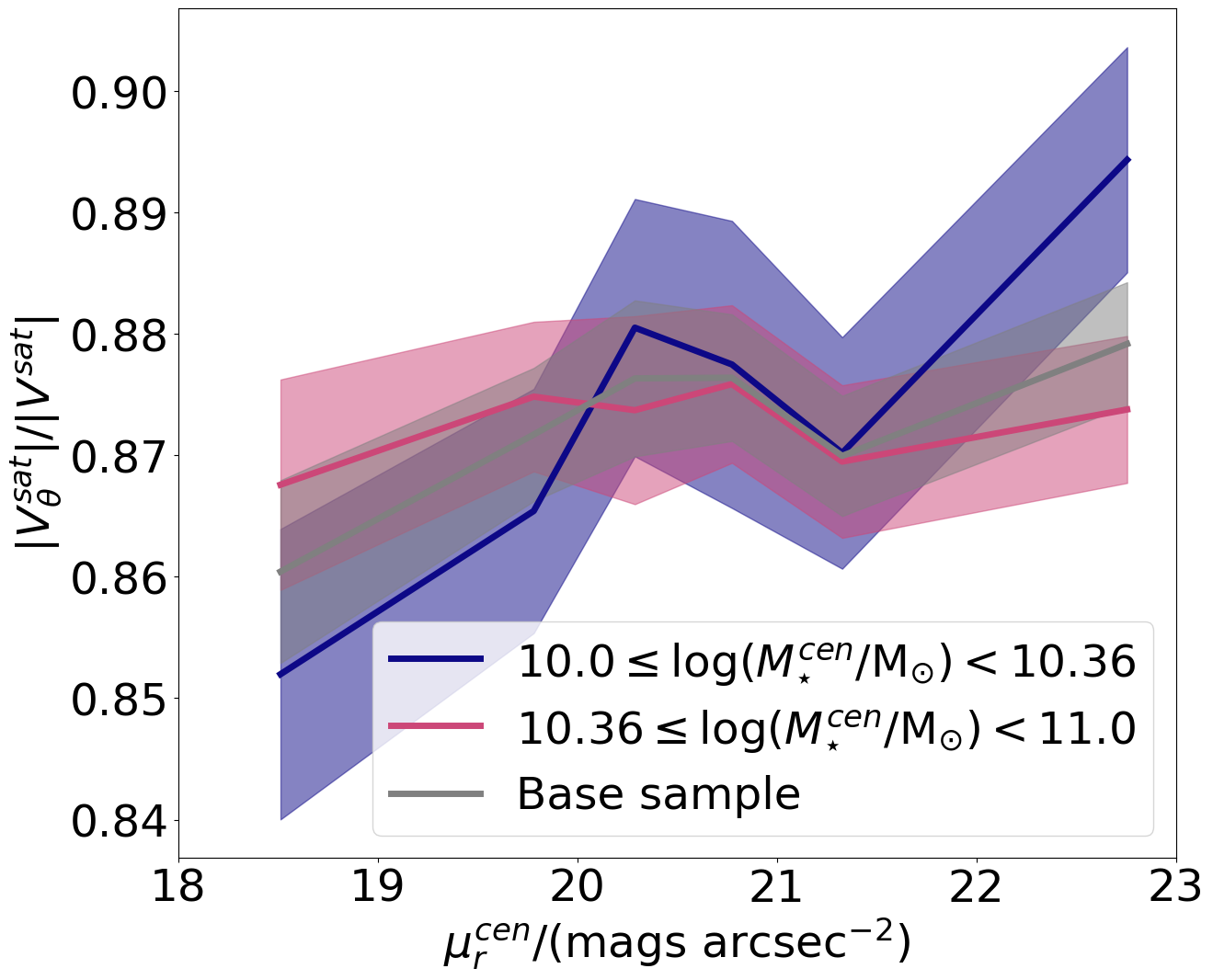}
    \includegraphics[width=0.475\textwidth]{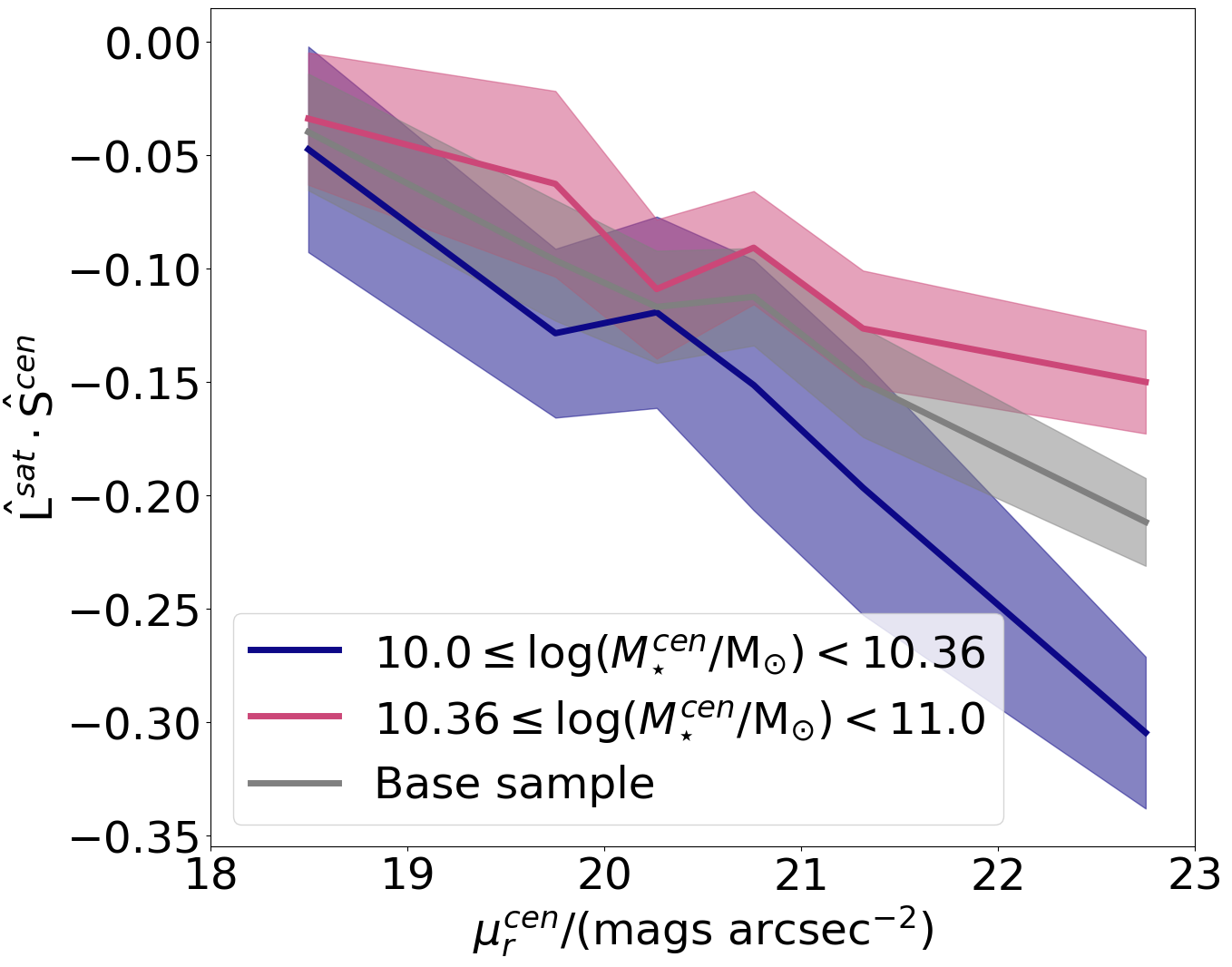}
    \caption{Magnitude of the tangential velocity normalized to the magnitude of the total velocity ($|V^{sat}_{\theta}|/|V^{sat}|$, left panel) and scalar product between the normalised angular momentum of the disc of the host central galaxy ($\hat{\mbox{S}}^{cen}$) and the normalised orbital angular momentum of each satellite ($\hat{\mbox{L}}^{sat}$, right panel). Both quantities are expressed as a function of $\mu^{cen}_r$, for the same stellar mass ranges as in Fig~\ref{fig:number_cont}. The solid lines correspond to the medians, and the shadowed region encloses the median bootstrap error.
    }
    \label{fig:vel_cont}
\end{figure*}

Another approach for the analysis of the central galaxy surface brightness relation to the local environment is provided in this section, where we determine the characteristics of satellites for central galaxies in the base sample. In this section we consider $\mu^{cen}_{r}$ as a continuous value instead of the previously defined LSBGs and comparison samples. We split the base sample into two sets of central galaxies according to their stellar mass, in low-mass $(10 \leq \log(M^{cen}_{\star}/\mbox{M}_{\sun}) <10.36)$, and high-mass $(10.36 \leq \log(M^{cen}_{\star}/\mbox{M}_{\sun})\leq 11.0)$ in order to have the same number of central galaxies, 1555 galaxies in each set. Then, we select the corresponding satellites following the same procedure detailed in Sect.~\ref{sec_satdef}. We obtain a total of 6212 satellites in the low-mass subsample, and 20040 satellites in the high-mass subsample.

Using the previously described set of satellites, Fig.~\ref{fig:number_cont} shows the median number of satellites as a function of $\mu^{cen}_{r}$ for the base sample (grey line), the high-mass (pink line) and, low-mass (indigo line) samples. We use the same colour code for all the remaining figures. In general, the number of satellites has a significant  host-surface-brightness dependence, where galaxies in both, the bright and faint end of the $\mu^{cen}_r$ distribution, host a larger number of satellites in comparison to  galaxies in the intermediate $\mu^{cen}_r$ range. At the same time, this signal has a stellar-mass dependence, being more evident in the range of massive galaxies (pink line). However, we notice the similarity between the relative increase in the number of satellites for the different central galaxy mass ranges. Overall, these results are consistent with Fig.~\ref{fig:numberSat}. 

Finally, in order to further explore the dynamics of satellites and taking into account the results of Fig.~\ref{fig:velocitydisp}, we have calculated the ratio of velocity modulus normal to the radial direction to the total velocity ($|V^{sat}_{\theta}|/|V^{sat}|$) for the satellite system as a function of $\mu^{cen}_{r}$ which is shown in the left panel of Fig.~\ref{fig:vel_cont} for the high-mass (pink), low-mass (indigo) and base (grey) samples. As it can be seen, there is an increasing trend for non-radial motions towards high values of $\mu^{cen}_{r}$. This is stronger for the low-mass central galaxies, presumably motivated by the excess of angular momentum of the dark matter haloes of LSBGs with respect to the comparison sample (see the rightmost panel in Fig~\ref{fig:derived_dist}). Studies like \citet{DiCintio2019, Wright2025, Wu2025} present evidence that LSBGs have undergone mergers in which the secondary orbit were, in general, coplanar respect to the primary gas disc. While, for HSBGs, this is not the case. The results shown here could be a sign of this behaviour.

We have also computed the scalar product between the normalised orbital angular momentum of the system of satellites ($\hat{\mbox{S}}^{sat}$) and the normalised angular momentum of the stellar component of the central galaxy ($\hat{\mbox{L}}^{cen}$) as a function of surface brightness $\mu^{cen}_{r}$. These results are in the right panel in Fig.~\ref{fig:vel_cont}, where a value of $\hat{\mbox{L}}^{sat} \cdot \hat{\mbox{S}}^{cen}>0$ indicates that the satellite system and the angular momentum of the central galaxy are co-rotating, while a negative value indicates that the galaxy and its satellites are counter-rotating. As seen in this figure, there is a strong anti-correlation of the alignment of the disc of the central galaxy and the orbit of the satellites with $\mu^{cen}_{r}$ which is more evident for the low-mass sample (indigo line). This means that the amount of orbital angular momentum in counter-rotating satellites with respect to the angular momentum of the disc decreases with the surface brightness of the central galaxy. We argue that this is an expected result in a scenario where the preferred accretion of co-rotating satellites onto galaxies evolving into LSBGs leaves their population of remaining satellites preferentially in counter-rotation. This last result is in agreement with the observational evidence that shows that the specific angular momentum of LSBGs tends to be higher when compared to that computed in HSGBs \citep{Perez-M2019, Jadhav2019}. This behaviour is also present in simulations \citep{Salinas2021, Perez-M2022, Zhu2023, Pallero2025}.

\section{Discussion}
\label{sec_disc}

The abundance of satellites and their dynamics can be used as a probe that could be crucial at understanding the emergence of low-surface-brightness galaxies. In this sense, this work can serve as a theoretical motivation for observational studies of these systems. This gains special interest considering the upcoming data from the Legacy Survey of Space and Time \citep[LSST,][]{Ivezic2019}, which will allow the study of currently undiscovered low-brightness regions. 
We argue that not only baryonic physics, but also gravity models may be tested in detail in future analysis of LSBGs studies. In the literature there are theoretical predictions about the effect of different gravity models on the formation of LSBGs \citep[for instance see][which analyses the effect of MOND models]{Muller2019b, Nagesh2024}. The remarkable difference in the spatial distribution and dynamics of satellites associated either to LSBGs or to the comparison sample in a $\Lambda$CDM scenario, which gives hints that these systems can provide relevant tests of these fundamental aspects.

Among the main results of our analysis is the higher cumulative number of satellites for LSBGs with respect to the comparison sample of similar stellar mass and morphology (Fig.~\ref{fig:numberSat}). We obtain an increasing difference in the abundance of satellites which reaches a factor $\sim 3$ at $R_{200}$. The possibility that this significant difference comes from a primordial origin or if it comes as a result of evolution was further explored. In particular, the cumulative distribution in both stellar and total mass was analysed for the LSBGs and comparison samples (Fig.~\ref{fig:massSat}). We found that at distances approaching $R_{200}$, both the total and stellar mass are similar in the two cases. Nevertheless, the mass of the satellite systems (stellar and total) at distances closer than $\sim 1/3~R_{200}$ is significantly larger for LSBGs than in the comparison sample. We argue that these points towards a dynamical evolution origin driven by a strong dynamical friction in systems evolving into HSBGs via mergers, which would cause a lack of massive satellites close to the central galaxies.

As with regards the dynamics of the satellite system, we find that their mean velocity dispersion as a function of the central stellar mass is highly correlated for the LSBGs. On the contrary, for satellites associated to the comparison sample, the correlation is significantly poorer (Fig.~\ref{fig:velocitydisp}). Since our LSBGs and comparison samples have a very similar halo mass, a different evolution of the system of satellites associated to the stellar density of the central galaxies must have been in action. This is further supported by the lack of corotating satellites in LSBGs with respect to the comparison sample as shown in Fig.~\ref{fig:vel_cont}. We claim that this is a clear hint of the accretion of corotating satellites by central galaxies in the process of a LSBG-type evolution. Accretions in co-rotation are expected to be less disruptive than for satellites counter-rotating or in radial trajectories. In fact, several recent studies indicate that the acquisition of angular momentum from accreted material, co-rotating or otherwise  aligned with the disc angular momentum, contributes to the formation of extended, diffuse discs by enhancing the net spin of the central system and limiting disc heating \citep{DiCintio2019,Wright2025,Wu2025}. Such aligned infall channels allow galaxies to maintain lower central stellar densities and more extended stellar distributions, consistent with the structural properties observed in LSBGs.
This orbital behaviour would be more frequently associated to the merger of satellites of the comparison sample, which would conserve the isotropic distribution of rotating and counter-rotating satellites, as it is observed in our analysis.

Our analysis of the satellite systems of galaxies with a range of surface brightness sheds light on the phenomena that take place in the formation of the structures with low surface brightness. The distribution and dynamics of satellites could provide evidence of the accretion history of central galaxies. The lack of previous hydrodynamical studies concerning the analysis of the satellite systems associated with LSBGs is related to the fact that the required resolution in hydrodynamical simulations has only being achieved in recent years with the advent of large hydrodynamical simulations like Illustris TNG \citep{Pillepich2018, Weinberger2017} or Eagle \citep{Crain2015, Schaye2015}. For instance, some of the studies on LSBGs that take advantage of these state-of-art simulations are \citet{Perez-M2022, Perez-M2024,Pallero2025}. In particular, \citet{Zhu2023} state that the population of LSBGs account only for 6\% of galaxies in the mass range given by $10^{10.2} \leq M_{\star}/\mbox{M}_{\sun} \leq 10^{11.6}$, challenging their observational detection. Despite these difficulties, some observational studies have been capable of characterizing the properties and environment of LSGBs \citep[for instance][]{Perez-M2019}. The predictions presented in this work could serve as inspiration for future observational analysis of LSB galaxies and their satellite systems. In particular, studies taking advantage of large upcoming surveys that will reach deeper magnitude limits than the available observational data, such as the LSST or Euclid \citep{Laureijs2011}.

\begin{acknowledgements}
The authors would like to thanks to the anonymous referee for their invaluable comments and suggestions, which help to improve this work greatly. SR and YY acknowledge financial support from FONCYT through the PICT 2019-1600 grant. JAB is grateful for partial financial support from the NSF-CAREER-1945310 and NSF-AST-2107993 grants. SP acknowledges partial support from ANPCyT (Argentina) through grant PICT 2020-00582. GG acknowledges the support of the ANID BASAL project CATA FB210003, as well as the support of the French-Chilean Laboratory for Astronomy (FCLA). 
\end{acknowledgements}

\bibliographystyle{aa}
\bibliography{biblio}
\end{document}